\documentclass[%
reprint,
amsmath,amssymb,
aps,
]{revtex4-1}

\usepackage[resetlabels]{multibib}
\usepackage{multirow}
\newcites{sec}{Reference}

\usepackage{graphicx}
\usepackage{dcolumn}
\usepackage{hyperref}
\hypersetup{colorlinks=true, citecolor=blue, urlcolor=blue, linkcolor=blue}
\usepackage{bm}
\usepackage{epstopdf}
\usepackage{titlesec}
\titleformat{\chapter}[display]
{\normalfont\Large\bfseries}{\thechapter}{11pt}{\Large}
\titleformat{\section}
{\normalfont\large\bfseries}{\thesection}{11pt}{\large}
\titlespacing*{\chapter}{0pt}{0pt}{15pt} 
\titlespacing*{\section}{0pt}{3.5ex plus 1ex minus .2ex}{2.3ex plus .2ex}
\usepackage[titletoc]{appendix}

\usepackage{algorithm}
\usepackage{algpseudocode}

\usepackage{enumitem}

\floatname{algorithm}{Algorithm}

\begin{document}

\preprint{APS/123-QED}

\title{Inferring Network Structures via Signal Lasso}
\thanks{Published in Phys. Rev. Research, 3, 043210 (2021)}

\author{Lei Shi$^{1,2}$}
\email{shi\_lei65@hotmail.com}
\author{Chen Shen$^1$}
\author{Libin Jin$^2$}
\author{Qi Shi$^3$}
\author{Zhen Wang$^3$}
\author{Stefano Boccaletti$^{4,5,7}$}

\affiliation{
	1. School of Statistics and Mathematics, Yunnan University of Finance and Economics, Kunming, 650221, China. \\
	2. Interdisciplinary Research Institute of Data Science, Shanghai Lixin University of Accounting and Finance, Shanghai 201209, China.\\
	3. Center for OPTical IMagery Analysis and Learning (OPTIMAL), Northwestern Polytechnical University, Xi'an 710072, China.\\
	4. CNR - Institute for Complex Systems, Via Madonna del Piano 10, 50019 Florence, Italy.\\
	5. Unmanned Systems Research Institute, Northwestern Polytechnical University, Xi'an, 710072, China.\\
	6. Moscow Institute of Physics and Technology, Institutskiy per., Dolgoprudny, Moscow Region, 141701 Russia.}
\date{\today}

\begin{abstract}
Inferring the connectivity structure of networked systems from data is an extremely important task in many areas of science. Most of real-world networks exhibit sparsely connected topologies, with links between nodes that in some cases may be even associated to a binary state (0 or 1, denoting respectively the absence or the existence of a connection). Such un-weighted topologies are elusive to classical reconstruction methods such as Lasso or Compressed Sensing techniques. We here introduce a novel approach called {\it signal Lasso}, where the estimation of the signal parameter is subjected to 0 or 1 values. The theoretical properties and algorithm of proposed method are studied in detail. Applications of the method are illustrated to an evolutionary game and synchronization dynamics in several synthetic and empirical networks, where we show that the novel strategy is reliable and robust, and outperform the classical approaches in terms of accuracy and mean square errors.
\end{abstract}

\keywords{Network reconstruction, Sparsity, Lasso, signal Lasso, evolutionary game, synchronization model}                       
\maketitle


\section{\label{sec1}Introduction}

Complex networks have a wide range of applications in various fields of science~\cite{strogatz2001exploring,albert2002statistical,watts2004small,erdHos1960evolution,watts1998collective,barabasi1999emergence,boccaletti2006complex}. In many cases, the pattern of the node-to-node interactions (the so called underlying network topology) is unknown and to be retrieved, and therefore uncovering the connectivity wiring of a distributed system's elementary units  based on observable quantities obtained from experiments is a very relevant issue. Network reconstruction is an inverse problem, and has received great attention recently~\cite{marsden1990network,butts2003network,squartini2017maximum,peixoto2018reconstructing,peixoto2019network,shandilya2011inferring}. Most existing approaches rely on the observation of the graph's dynamics (for instance that exhibited by networked oscillators) as the result of continuous differential equations. However, gathering a continuous-in-time information on the dynamical evolution of the graph is not always possible in social, economic, and even biological systems. Evolutionary-game based dynamics, for instance, characterizes many relevant situations, and is intrinsically discrete in time. In such a latter case, the problem can be transformed into a statistical linear model, with sparse and high dimensional properties. The Compressed Sensing ($CS$) and the Lasso method have been proposed for the purpose of network reconstruction~\cite{han2015robust,wang2011network}. These two methods provide a good performance in uncovering the network topology, even for cases in which only a limited time series is observed, and the Lasso method is furthermore quite robust against various sources of noise which may affect the data.

Reconstructing the connectivity pattern of un-weighted networks implies, however, estimating connection parameters between nodes which have true values of either 0 or 1. All existing methods (including $CS$ or Lasso) have the ability of shrinking the estimates to zero when networks are sparse, but fail in compressing the estimates to 1, and this latter fact unavoidably reduces their estimation accuracies. In this Letter we proposed a novel method, called {\it signal Lasso}, able to uncover the graph's topology with a small amount of data. The main idea behind the new method is to add a control term of L1-norm in the penalty function of Lasso, able to shrink the estimates to 1 and therefore overcoming the above limitation. The result is an estimate of the signal parameters with values exactly equal to 0 or 1, for non-existent or existent links between nodes, respectively. We will show that the new method is robust (at least as Lasso) against noise and missing nodes, due to the fact that a least square error control term is included. Validation and assessment of the method's robustness is illustrated with reference to spatial ultimatum game and Kuramoto model in synchronization dynamic, for three typical classes of synthetic networks (Erd\"os-R\'enyi (ER) random ~\cite{erdHos1960evolution}, Watts-Strogatz (WS) small word ~\cite{watts1998collective} and Barab\'asi-Albert (BA) scale free ~\cite{barabasi1999emergence} networks) as well as for some empirical networks. High prediction accuracies are obtained also for extremely small data sets, as compared with $CS$ and Lasso methods. Therefore, our results are of potential interest for many applications in social, economic, and biological systems, in all cases in which recovering hidden networks from very limited samples of data is required.

\begin{figure}
	\centering{\includegraphics[width=0.46\textwidth]{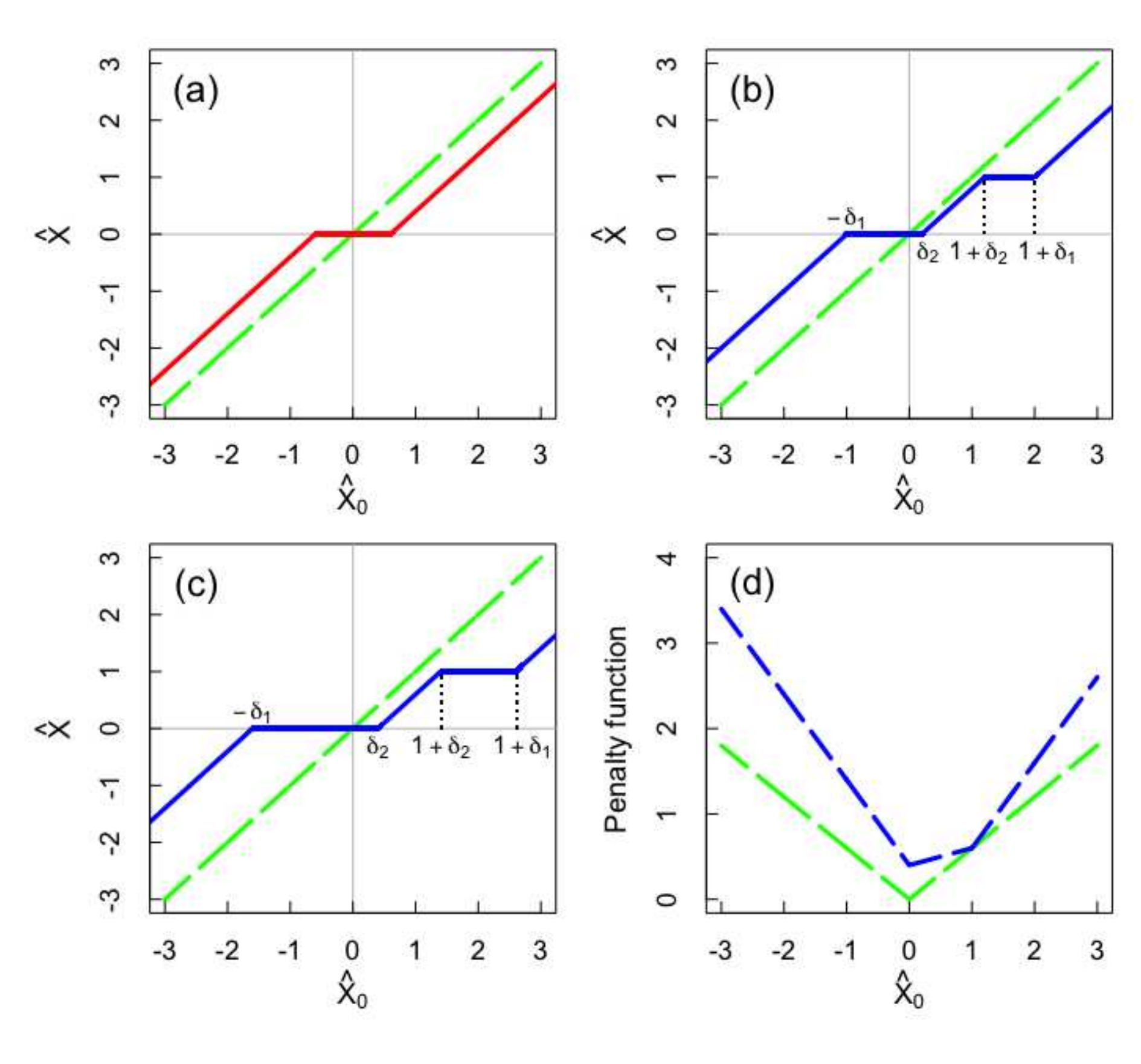}}
	\caption{(a) Solution $\hat X$ (red curve) for the Lasso method ($\lambda=0.6$) vs. the solution $\hat X_0$ obtained under the least square; (b) solution $\hat X$ (blue curve) for the signal Lasso method ($\delta_1=1, \delta_2=0.2$) vs. $\hat X_0$; (c) same as in panel (b), but for $\delta_1=2$ and $\delta_2=0.4$. In all three panels, we draw a green line $\hat X=\hat X_0$ for reference.  (d) Penalty function (see text for definition) of the signal Lasso method (blue line) for $\delta_1=1, \delta_2=0.2$, and of the Lasso method (green line) for $\lambda=0.6$.}
	\label{fig:1}       
\end{figure}

The problem under study can be framed as the solution of the following linear equation ~\cite{han2015robust}:
\begin{equation}\label{eq1}
Y=\Phi X,
\end{equation}
where the $n\times p$ matrix $\Phi$ and the $n\times 1$ vector $Y$ are accessible, whereas $X$ is a parameter vector which is related to the adjacency matrix $[a_{ij}]$ of the network (having elements $a_{ij}=1$ for all pairs $i,j$ of connected nodes, and $a_{ij}=0$ otherwise). Therefore, $X$ is a signal parameter with values of 1 or 0, and the purpose of network reconstruction is to estimate $X$ from $Y$ and $\Phi$. Normally, the dimension of $X$ is much higher than the dimension of $Y$, and the problem displays a sparsity property due to the fact that only a few non zero elements are present in $X$~\cite{boccaletti2006complex,friedman2010regularization}. Recently, the Lasso method was proposed to find a solution of Eq. (\ref{eq1}) from data, which is obtained by solving a convex optimization problem in view of a variable selection~\cite{tibshirani1996regression,hastie2015book}. However, the Lasso method is able to reduce the parameter estimates of unimportant predictors in $X$ to 0, but the estimators of non-zero elements fail to be conducted towards their {\it real} values 1. This latter feature of the methods causes the estimators of parameters with value 1 to have a rather low accuracy, as we will show along the present Letter. Other methods, such as SCAD~\cite{fan2009network}, adaptive Lasso~\cite{zou2006adaptive}, group Lasso~\cite{bach2008consistency}, or elastic net~\cite{zou2005regularization} and compressed sensing (CS)~\cite{donoho2006compressed}, which intrinsically focus on zero elements suffer the same restriction as Lasso, and therefore fail to give accurate descriptions of the non-zero elements of $X$.

\section{\label{sec2}Signal Lasso}

\begin{figure}
	\centering{\includegraphics[width=0.45\textwidth]{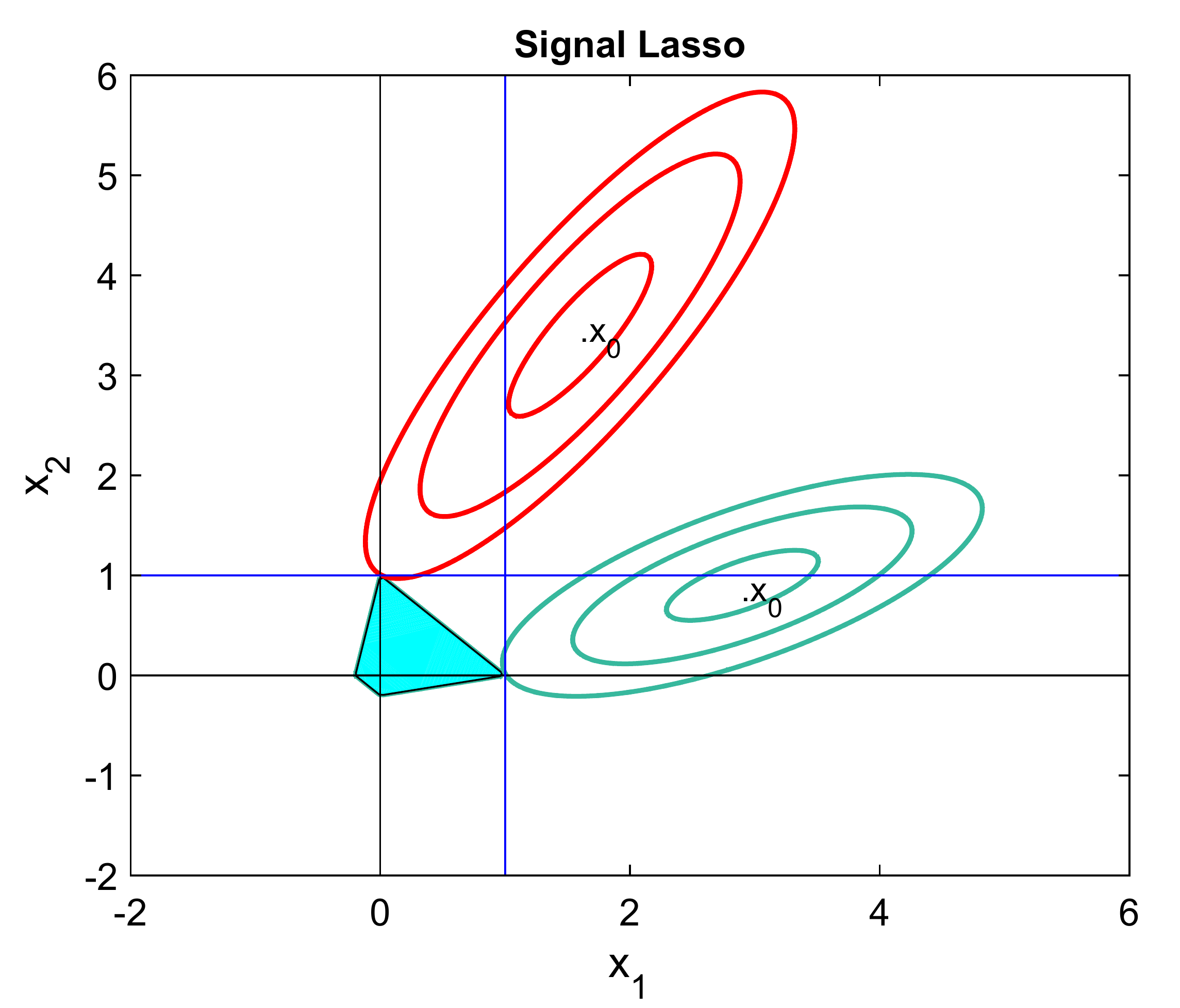}}
	\caption{Sketch of how estimation is obtained in the signal Lasso method. The solid blue area is the constraint region $\lambda_1 \sum_{j=1}^2 \vert x_j\vert +\lambda_2 \sum_{j=1}^2 \vert x_j-1\vert \le 1$, with $\lambda_1=0.6, \lambda_2=0.4$, while the ellipses are the contours of the residual sum of squares in two situations, centered around the solutions $X_0$ under the least square. In both cases, the method finds the first point where the elliptical contours hit the corner of the constraint region, which corresponding to (0,1) for upper ellipses or (1,0) for lower ellipses. Interested readers can compare this scenario with the case of the Lasso method (reported in Ref. ~\cite{hastie2015book}).}
	\label{fig:2}       
\end{figure}

We exclude an intercept in model (\ref{eq1}) to consist with the model we studied in this paper. The least square method
for estimating $X$ is found by minimizing $\sum_{i=1}^n (y_i-\phi'_{i} X)^2$, while Lasso method~\cite{tibshirani1996regression} subject
a constraint that $\sum_{i=1}^n \vert X_i \vert <t$, where $t >0$ is a tuning parameter, $\phi_i$ denote the $i$th row of $\Phi$. The solution of Lasso can be found by  minimizing following penalty likelihood
\begin{equation}\label{eq2}
\frac{1}{2}\sum_{i=1}^n (y_i-\phi'_{i} X)^2+\lambda\sum_{j=1}^p\vert X_j\vert
\end{equation}
with regulation parameter $\lambda$. However in our problem, $X_j$ must be either 0 or 1, therefore a novel method, called {\it signal Lasso} is proposed by minimizing following function
\begin{equation}\label{eq3}
\frac{1}{2}\sum_{i=1}^n (y_i-\phi'_{i} X)^2+\lambda [(1-\alpha)\sum_{j=1}^p\vert X_j\vert + \alpha\sum_{j=1}^p\vert X_j -1\vert],
\end{equation}
where $\bf{1}_p$ is a $p\times 1$ vector with all elements equal to 1, $\Vert X\Vert_k$ is the $k$-norm of vector X defined by $\Vert X\Vert_k=(\sum_{j=1}^p \vert X_j\vert ^k)^{1/k}$, $\lambda > 0$ and $\alpha \in (0,1)$ are two tuning parameters. We add term of $\sum_{j=1}^p\vert X_j -1\vert$ in the penalty term since some elements of $X$ should be 1. This is a compromise between the terms shrinking to 0 and 1 and we hope some elements of $X$ will close to 0 and others to 1. Minimization of Eq. (\ref{eq3}) can lead to the solution of optimal estimation of $X$. In general, it is impossible to make $X_j$ close to 0 and 1 at the same time, a common way in computational mathematics is to give a probability such as $\alpha$ to signal group and $1-\alpha$ to non-signal group as shown in Eq. (\ref{eq3}), the value of $\alpha$ can be determined by the data set at hand.

\begin{figure}
	\centering{\includegraphics[width=0.47\textwidth]{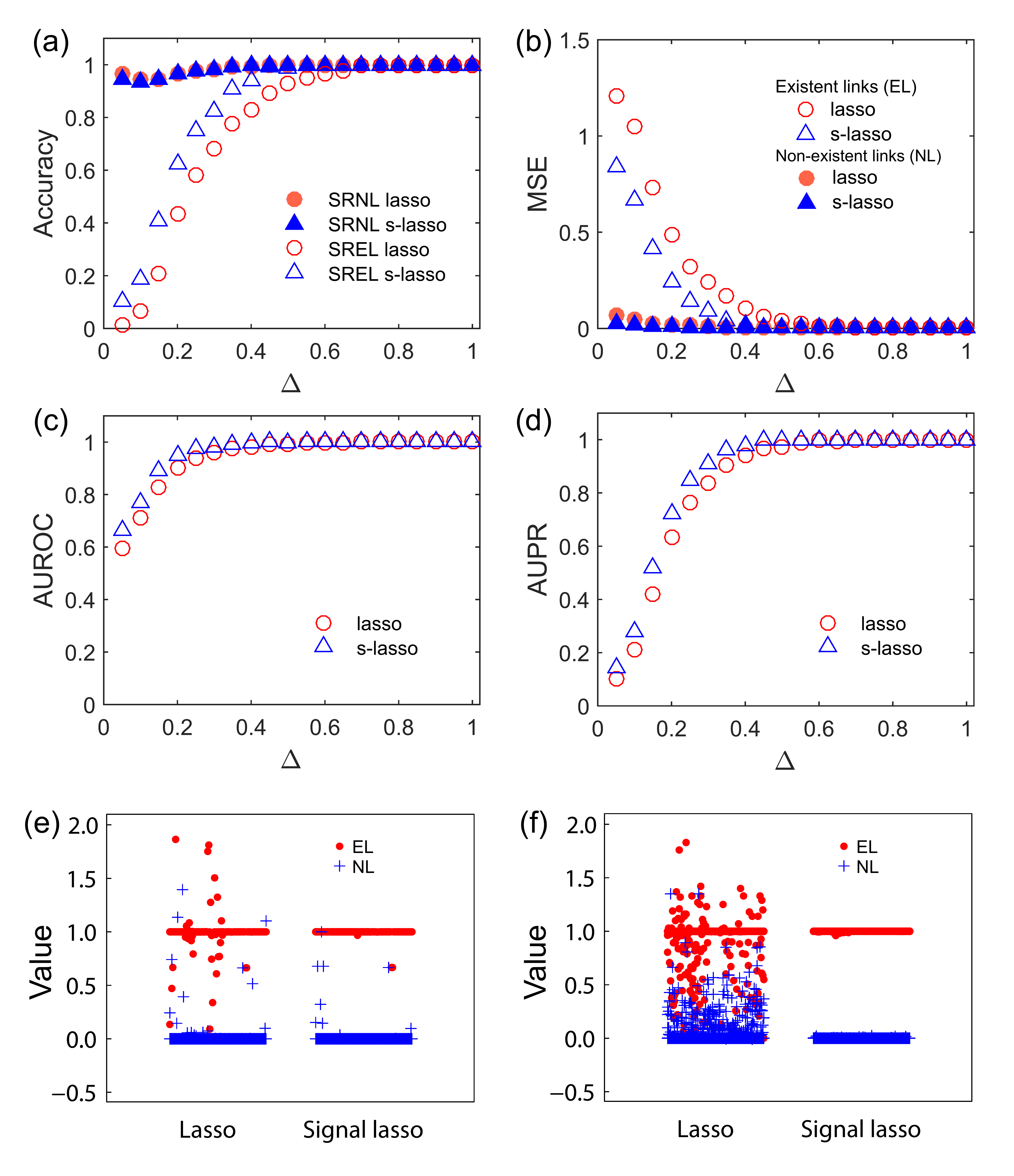}}
	\caption{Accuracy in the reconstruction vs. $\Delta=L/N$, attained by  the Lasso and signal Lasso methods. Panels (a-d) refer to a $SUG$ played on top of Barab\'asi-Albert (BA) scale-free networks, and report (a) the SREL and SRNL criteria, (b) the MSE criterion, (c) the AUROC criterion, and (d) the AUPR criterion (see main text for all definitions). (e-f) Reconstructed values of the elements of the vector $X$ obtained by the Lasso and signal Lasso methods for a $SUG$ played on top of the karate club network with $\Delta=0.5$ (e) and on Erd\"os-R\'enyi (ER) random networks with average degree 12 for $\Delta=0.4$ (f). $N$ is equal to 100 and 34 for BA and karate club networks, respectively.}
	\label{fig:3}       
\end{figure}

Eq. (\ref{eq3}) can be rewritten as a more general form by
\begin{equation}\label{eq4}
\min_{X}\left\{\frac{1}{2} \left\|Y-\Phi X\right\|^2_2 +\lambda_1  \left\|X\right\|_1+  \lambda_2  \left\|X-\mathbf{1}_p\right\|_1 \right\},
\end{equation}
where $\bf{1}_p$ is the $p\times 1$ vector with all elements equal to 1, $\lambda_1$  and $\lambda_2$ are  non-negative regularization parameters and $\left\|X\right\|_k$ is the $k-$ norm of the vector $X$. Furthermore, the presence of a least square term $\left\|Y-\Phi X\right\|^2_2$ in Eq. (\ref{eq4}) makes the solution robust against noise. Since Eq. (\ref{eq4}) is a convex optimization problem, the solution is unique and some algorithms can be developed to solve it ~\cite{friedman2010regularization,john2010elements}. Moreover, it has to be noticed that  Eq. (\ref{eq4}) reduces to the classical Lasso method for $\lambda_2=0$.

Let us now focus on the case in which the columns of $\Phi$ are orthogonal, with $\Phi'\Phi=I_p$ ($I_p$ denoting the identity matrix of order $p$), and let us have $\delta_1=\lambda_1+\lambda_2$ and $\delta_2=\lambda_1-\lambda_2$. Then, for $k=1,\cdots,p$, the solution has the following form [see the Appendix for full details]:
\begin{equation}\label{eq5}
\hat X_k=\left\{
\begin{array}{l}
(\hat X_{k0}+\delta_1)_{-}, \ \ \ \ \ \ \ \hat X_{k0}\leq0, \\
(\hat X_{k0}-\delta_2)_{+}, \ \ \ \ \ \ \ 0<\hat X_{k0}\leq 1+\delta_2, \\
\max\{1, \hat X_{k0}-\delta_1\}, \ \ \hat X_{j0}> 1+\delta_2,
\end{array}\right.
\end{equation}
where $\hat X_{k0}$ and $\hat X_k$ are the $k$th element of $\hat X_{0}$ and $\hat X$, respectively, and $(B)_{+}$ [$(B)_{-}$] in Eq.~(\ref{eq5}) denotes the positive (negative) part of the vector $B$, meaning that $(B)_{+}=B$ if $B\ge 0$ [$(B)_{-}=B$ if $B\le 0$] and $(B)_{+}=0$ if $B < 0$ [$(B)_{-}=0$ if $B > 0$]. Notice furthermore that there is a one-by-one transformation between $(\lambda_1,\lambda_2)$ and $(\delta_1,\delta_2)$. All details on the theoretical basis and algorithm of our method are given in Appendix.

In Fig. \ref{fig:1} we report the solution $\hat X$ as a function of $\hat X_0$ for the Lasso [panel a)] and for the signal Lasso method [panels b) and c)]. From panel a) of Fig. \ref{fig:1} one immediately sees that the Lasso method is able to shrink the parameter estimator $\hat X$ to zero for $\vert \hat X_0 \vert \le 0.6$: the solution in this case is given by $\hat X_k=sgn(\hat X_{k0})(\vert \hat X_{k0}\vert -\lambda)_{+}$, where $\lambda=0.6$, sgn(z) is a threshold function with sgn(z)=sign(z) if $z\neq 0$ and some value lying in [-1, 1]. However, for $\vert \hat X_{k0} \vert > 0.6$, the classical Lasso method is only able to translate the values by a constant factor.
Panel b) of Fig. \ref{fig:1} reports the results of the signal Lasso method, for $\lambda_1=0.6$ and $\lambda_2=0.4$ (this choice is equivalent to $\delta_1=1.0, \delta_2=0.2$).
Some geometry properties of the signal Lasso method are apparent: values in the interval $-\delta_1\le\hat X_{k0}\le \delta_2$ are shrunk to $\hat X=0$, and values satisfying $1+\delta_2\le\hat X_{k0}\le 1+\delta_1$ are shrunk to 1. Other values are translated by a constant. Therefore, the novel method is also able to compress larger values of the parameter estimator to 1, resulting in a improvement of the estimation accuracy. Panel c) of Fig. \ref{fig:1} reports the results for the case $\delta_1=1.8, \delta_2=0.4$, in which the length of intervals shrunk to zero and 1 are increased. Finally, in panel d) of Fig. \ref{fig:1} we report the penalty function
$PF=\lambda_1  \left\|X\right\|_1+  \lambda_2  \left\|X-\mathbf{1}_p\right\|_1$
vs. $X$, for the same conditions as in Fig. \ref{fig:1}(b), in a way that the shrinking directions toward 0 and 1 are visible.
The constraint region in the signal Lasso method is a polygon with corners from 4 to eight (see SI for full details). Fig.\ref{fig:2} shows the estimation picture, where the solution [such as (0,1) or (1,0) in a two dimensional case] is found from the elliptical residual sum of the square hit with the corner of the constraint region.
When the columns of $\Phi$ are not orthogonal and $p$ is sufficiently larger than $n$, the solution can be found by convex optimization algorithms. The choice of $\lambda_1$ and $\lambda_2$ (or equivalently, of $\delta_1$ and $\delta_2$) is guided by cross validation methods (see Appendix for full details).

\begin{table*}
	\centering
	\caption{Maximum value of the data ratio $\Delta$ which simultaneously ensures the value of at least 0.05 for the accuracy in reconstructing existent and non-existent links. The methods compared are  the compressing sensing (CS), the Lass and the signal Lasso. The comparison is made with SUG played on scale-free networks (BA), small-world Watts-Strogatz networks, and random ER graphs. All results refer to averages over ten independent realizations. N is the network size, $\left \langle k \right \rangle$ the average degree of the  network, and $\sigma$ the amplitude of the Gaussian noise. }
	\label{t1}
	\setlength{\tabcolsep}{7mm}{
		\begin{tabular}{cccccc} \toprule
			&  &   & SF & WS & ER \\
			N &  $\left \langle k \right \rangle$ &  $\sigma$  &  & CS/lasso/signal lasso &  \\
			\hline
			100 & 6 & 0 & 0.44/0.44/\textbf{0.34} & 0.26/0.26/\textbf{0.24} & 0.29/0.29/\textbf{0.27} \\
			100 & 6 & 0.05 & nan/0.49/\textbf{0.4} & nan/0.36/\textbf{0.33} & nan/0.34/0.34 \\
			100 & 6 & 0.3 & nan/1.5/\textbf{1.4} & nan/1.2/1.2 & nan/1.4/1.4 \\
			100 & 12 & 0 & 0.55/0.55/\textbf{0.4} & 0.39/0.39/\textbf{0.36} & 0.39/0.39/\textbf{0.36} \\
			100 & 18 & 0 & 0.60/0.60/\textbf{0.42} & 0.43/0.43/\textbf{0.39} & 0.46/0.46/\textbf{0.41} \\
			100 & 30 & 0 & 0.68/0.68/\textbf{0.46} & 0.55/0.55/\textbf{0.49} & 0.59/0.59/\textbf{0.5} \\
			\hline
	\end{tabular}}
\end{table*}

\section{\label{sec3}Reconstruction of connectivity based on evolutionary game}

In what follows, we apply the method to reconstruct the connectivity pattern of players of the spatial ultimatum game ($SUG$). In such a framework, each node of a network of size $N$ represents a player, and links of the graph stand for game relationships between players. In the SUG~\cite{szolnoki2012defense,fehr2003nature}, each player $i$ is assigned an arbitrary strategy pair ($p_i$, $q_i$), with $p_i$, $q_i$ $\in$ [0, 1] which are drawn from a uniform distribution and which represent its offering and acceptance level, respectively. The payoff $P_{ij}$ that player $i$ obtains from playing with player $j$ is calculated as follows:
\begin{equation}\label{eq8}
P_{ij}=\left\{
\begin{array}{l}
p_j+1-p_i, \quad p_i\geq q_j,\quad p_j\geq q_i,  \\
1-p_i, \qquad \quad p_i\geq q_j,\quad p_j < q_i,  \\
p_j, \qquad \quad \quad \ \  p_i < q_j,\quad p_j\geq q_i,  \\
0,   \qquad \qquad \quad  p_i < q_j,\quad p_j < q_i,  \\
\end{array}\right.
\end{equation}
Then, the total payoff $F_i$ is calculated as $F_i=\sum_{j} a_{ij} P_{ij}$, where $a_{ij}=1$ if connected and 0 otherwise. At each round of the game, each player first calculates its total payoff and then imitates with a certain probability the two strategies (p, q) of a randomly selected player in its direct neighborhood. Precisely, player $x$ adopts the strategy of player $y$ with probability $W=1/\{1+\exp{[(F_x-F_y)/K]}\}$~\cite{szabo1998evolutionary,szabo2005phase}. In order to make the model more realistic, we also account for mutation: at each round of the game, all players adjust their strategy (p, q) according to $(p_i(t+1), q_i(t+1))=(p_i(t)+\delta, q_i(t)+\delta)$, where $\delta \in [-\epsilon, \epsilon]$ is a small random number (in our examples $\epsilon =0.05$).

Now $F_i=\sum_{j=1, j \neq i}^N a_{ij} P_{ij}$ can be written as a linear regression model
\begin{equation}\label{eq9}
Y_i=\Phi_i X_i,
\end{equation}
where $Y_i=(F_i(t_1), F_i(t_2), \cdots, F_i(t_L))'$, $X_i=(a_{i1}, \cdots, a_{iN})'$, and
$\Phi_i$  has the form of
\begin{displaymath}
\left(\begin{array}{cccccc}
P_{i1}(t_1) & \cdots & P_{i,i-1}(t_1) & P_{i,i+1}(t_1) & \cdots & P_{i_N}(t_1)\\
P_{i1}(t_2) & \cdots & P_{i,i-1}(t_2) & P_{i,i+1}(t_2) & \cdots & P_{i_N}(t_2)\\
\vdots & \vdots & \vdots & \vdots  & \vdots & \vdots   \\
P_{i1}(t_L) & \cdots & P_{i,i-1}(t_L) & P_{i,i+1}(t_L) & \cdots & P_{i_N}(t_L)
\end{array}\right),
\end{displaymath}
Let $Y=\Phi X$, where $Y=(Y'_1, \cdots, Y'_N)'$, $X=(X'_1, \cdots, X'_N)'$, $\Phi=diag(\Phi_1, \Phi_2, \cdots, \Phi_N)$, then Eq.(\ref{eq9}) can be converted to general form of Eq. (\ref{eq1}).
We here suppose that all total payoffs are accessible at $L$ successive times, so that one can write a linear regression model of the form of Eq.~(\ref{eq1}), where the exact form of $Y$ and $\Phi $ are given in Appendix.

In order to quantify the performance of the method we use the mean squared error ($MSE$) to measure how close the inferred adjacency matrix $A^\star$ is to the true matrix $A$. Specifically, we use the following equations for existent and non-existent links:
\begin{equation}
MSE_k=\frac{\sum_{i}^N\sum_{j\in \Omega_k} (a_{ij}-a^\star_{ij})^2}{N_k}, k=1, 2,
\end{equation}
where $\Omega_1$ ($\Omega_2$) is the set of existent (non-existent) links, and $a_{ij}^\star$  ($a_{ij}$) are the elements of $A^\star$ ($A$). For a complete comparison, we compute the area under the receiver operating characteristic curve (AUROC) and the area under the precision recall curve (AUPR) as defined in Ref.~\cite{han2015robust}, as well as the success rates for detection of existing links (SREL) and non-existing links (SRNL), as defined in Ref.~\cite{wang2011network}.

\begin{table}
	\centering
	\caption{Same as in Table ~\ref{t1}, but here SUG was played on some real networks~\cite{zachary1977information, lusseau2003bottlenose, girvan2002community}. All results are averages over ten independent realizations. N is the network size, $\left \langle k \right \rangle$ is the average degree of the given network.}
	\label{t2}
	\setlength{\tabcolsep}{3mm}{
		\begin{tabular}{cccccc} \toprule
			Networks & N & $\left \langle k \right \rangle$ & CS & lasso & signal lasso \\
			\hline
			Karate & 34 & 4.6 & 0.62 & 0.62 & \textbf{0.45}\\
			Dolphins & 62 & 5.1 & 0.38 & 0.38 & \textbf{0.34} \\
			Football & 115 & 10.7 & 0.29 & 0.29 & \textbf{0.27} \\
			\hline
	\end{tabular}}
\end{table}

Fig. \ref{fig:3} reports the comparison between the accuracy attained by signal Lasso and Lasso methods for Barab\'asi-Albert (BA) scale-free networks with $N=100$ when the criteria of SREL and SRNL (panel (a)), of MSE (panel (b)), of AUROC (panel (c)) and of AUPR (panel (d)) are adopted. The amount of data is scaled as $\Delta=L/N$. One clearly sees that the signal Lasso method improves always the reconstruction accuracy (as compared with the Lasso method), no matter which comparison criterion is adopted. For instance (and as an example), at $\Delta=0.4$ one has $MSE=0.02$ for signal Lasso and $MSE=0.1$ for Lasso. We also used the karate club network~\cite{zachary1977information} and ER networks with average degree $\left \langle k \right \rangle=12$. The results are shown in Fig. \ref{fig:3}(e) and Fig. \ref{fig:3}(f), respectively. As predicted, the signal Lasso method effectively shrink the parameter values of $X$ to 0 or 1 in both cases, while the performance of the Lasso method is poorer. A more complete comparison is reported in Table ~\ref{t1}, where we confront the novel method with Lasso and compressing sensing (CS) in terms of the minimal number of data (the maximum $\Delta$) needed to achieve an accuracy value of at least 0.05 in both $MSE_1$ (associated to the reconstruction task of existent links) and $MSE_2$ (associated to the reconstruction task of non-existent links), for different types of networks. The performance of signal Lasso  is always better than the other two methods, and the advantage is more prominent for networks with higher average degree. When a Gaussian noise  is present, the advantage is not so big, and yet the highest reconstruction accuracy is obtained for the signal Lasso method. In Table ~\ref{t2} the same comparison is done for some real-world networks, with similar and consistent results. For more details on these latter results, we address the interested reader to the Supplementary Materials.

\section{\label{sec4} Reconstruction of connectivity based on synchronization model}

\begin{figure*}[htbp!]
	\centering{\includegraphics[width=0.99\textwidth]{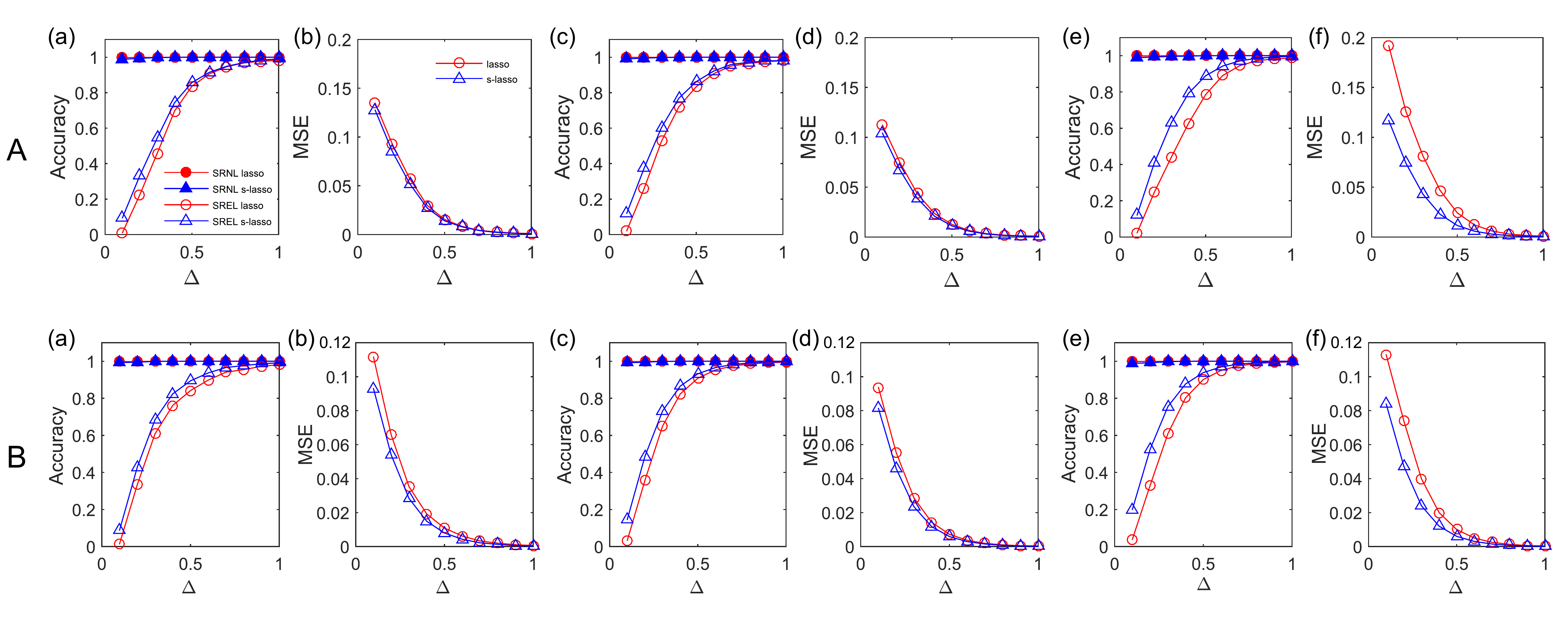}}
	\caption{Accuracy in the reconstruction vs. $\Delta=L/N$, for Kuramoto model attained by  the Lasso and signal Lasso methods. Panels (a-b) refer to Erd\"os-R\'enyi (ER) random networks with average degree 6. panels (c-d) refer to Barab\'asi-Albert (BA) scale-free networks, and panel (e-f) refer to small world (WS) network. We report (a,c,e) the SREL and SRNL criteria, (b,d,f) the MSE criteree, where  panel
		A for network size $N=30, c=10$ and panel B for network size $N=60, c=10$. Each point is averaged over 50 simulations. }
	\label{fig:4}    
\end{figure*}

Second we study the reconstruction of network in synchronization model ~\cite{boccaletti2014structure,timme2007revealing,Wu2012inferring}. We employ the simple Kuramoto model for illustration which has the following governing equation:
\begin{equation}\label{eq10}
\frac{d\theta_i}{dt}=\omega_i +c\sum_{j=1}^N a_{ij} sin(\theta_j -\theta_i),
\end{equation}
$i=1,\cdots, N$, where the system is composed of $N$ oscillators with phase $\theta_i$ and coupling strength $c$, each of the oscillators has its own intrinsic natural frequency $\omega_i$, $a_{ij}$ is the adjacency matrix of a give network and is need to be estimated in network reconstruction. The method proposed here can be modified to give solution but with a little difference with Eq. (\ref{eq1}) since there is a heterogeneity parameter $\omega_i$.

To solve above equation, the Euler method was employed to generate time series with an equal time step $h$. Let time increment of $\theta$ are used to approximate the $\dot\theta={d\theta_i}/{dt}$, in another word $\dot\theta=[\theta(t+h)-\theta(t)]/h$,  then we have following update formula
$$
\frac{[\theta_i(t+h)-\theta_i(t)]}{h}=\omega_i+c\sum_{j=1}^N a_{ij} * sin(\theta_j (t)-\theta_i(t)).
$$
Once initial values of $\theta_i(0)$ for $i=1, \cdots, N$ are given, the estimator of $\theta_i$  for $i=1, \cdots, N$ can be obtained iteratively, and a sequence of time series data are presented. Let $Y_i=(y_{i1}, \cdots, y_{iL})'$, $y_{it}=[\theta_i(t+h)-\theta_i(t)]/h$,
\begin{displaymath}
\Phi_i=
\left(\begin{array}{cccc}
\phi_{i1}(1) & \phi_{i,2}(1) & \cdots &  \phi_{i_N}(1)\\
\phi_{i1}(2) & \phi_{i,2}(2) & \cdots &  \phi_{i_N}(2)\\
\vdots & \vdots & \vdots & \vdots      \\
\phi_{i1}(L) & \phi_{i,2}(L) & \cdots &  \phi_{i_N}(L)\\
\end{array}\right),
\end{displaymath}
where $\phi_{ij}(t)=c \times sin(\theta_j (t)-\theta_i(t))$, $X_i=(a_{i1}, \cdots, a_{iN})'$, then
\begin{equation}\label{eq13}
Y_i=\omega_i {\bf 1}_L+\Phi_i X_i,
\end{equation}
where ${\bf 1}_L$ denote a $L\times 1$ vector with all element 1.  The Eq. (\ref{eq13}) can be stacked as a linear model as $Y=Z\omega+\Phi X$,  where $Y=(Y'_1, \cdots, Y'_N)'$, $\Phi$=diag($\Phi_1, \Phi_2, \cdots, \Phi_N)$, $Z=I_N \otimes {\bf 1}_L$, where $\otimes$ denote the Kronecker product, $I_N$ is a identity matrix with order $N$.
The method proposed here can be modified to give solution but with a little difference with Eq. (\ref{eq4}) since there is a heterogeneity parameter $\omega_i$. Solution of $X$ and $\omega$ using signal Lasso method can be obtained by minimizing

\begin{equation}\label{eq15}
\min_{X,\omega}\left\{\frac{1}{2} \left\|Y-Z\omega-\Phi X\right\|^2_2 +\lambda_1  \left\|X\right\|_1+  \lambda_2  \left\|X-\mathbf{1}_p\right\|_1  \right\}.
\end{equation}
Note that parameter $\omega$ is not shrunk in minimization but need to be estimated as an intrinsic natural frequency. The algorithm can be modified slightly as that in Section ~\ref{sec2}.


We generate time series using Euler method from Eq.(\ref{eq10}) with equal time step 0.01, where $\omega_i$ is generated from a uniform random variable in $[-0.5,0.5]$.
Fig. \ref{fig:4} give the results of reconstruction accuracy (SREL, SRNL and MSE) in ER, WS and BA networks respectively for $N=30, 60$ and coupling strength c=10 (see Ref.~\cite{Wu2012inferring} for similar setting in this model), and the results clearly shown signal Lasso method is outperformed to Lasso methods.

The scenarios with noise in this model can be similarly conducted, for example we can generate the data from a normal distribution with mean zero and variance $\sigma^2$ and calculate the  reconstruction accuracy (SREL, SRNL and MSE) in terms of ER, WS and BA networks, respectively.
We find that the reconstruction are more difficult with noise than that without noise, however they all show that the signal Lasso method is superior to Lasso method.

\section{\label{sec5}{Reconstruction of connectivity based on human behavior experiment}}

\begin{figure*}[htbp!]
	\centering{\includegraphics[width=0.99\textwidth]{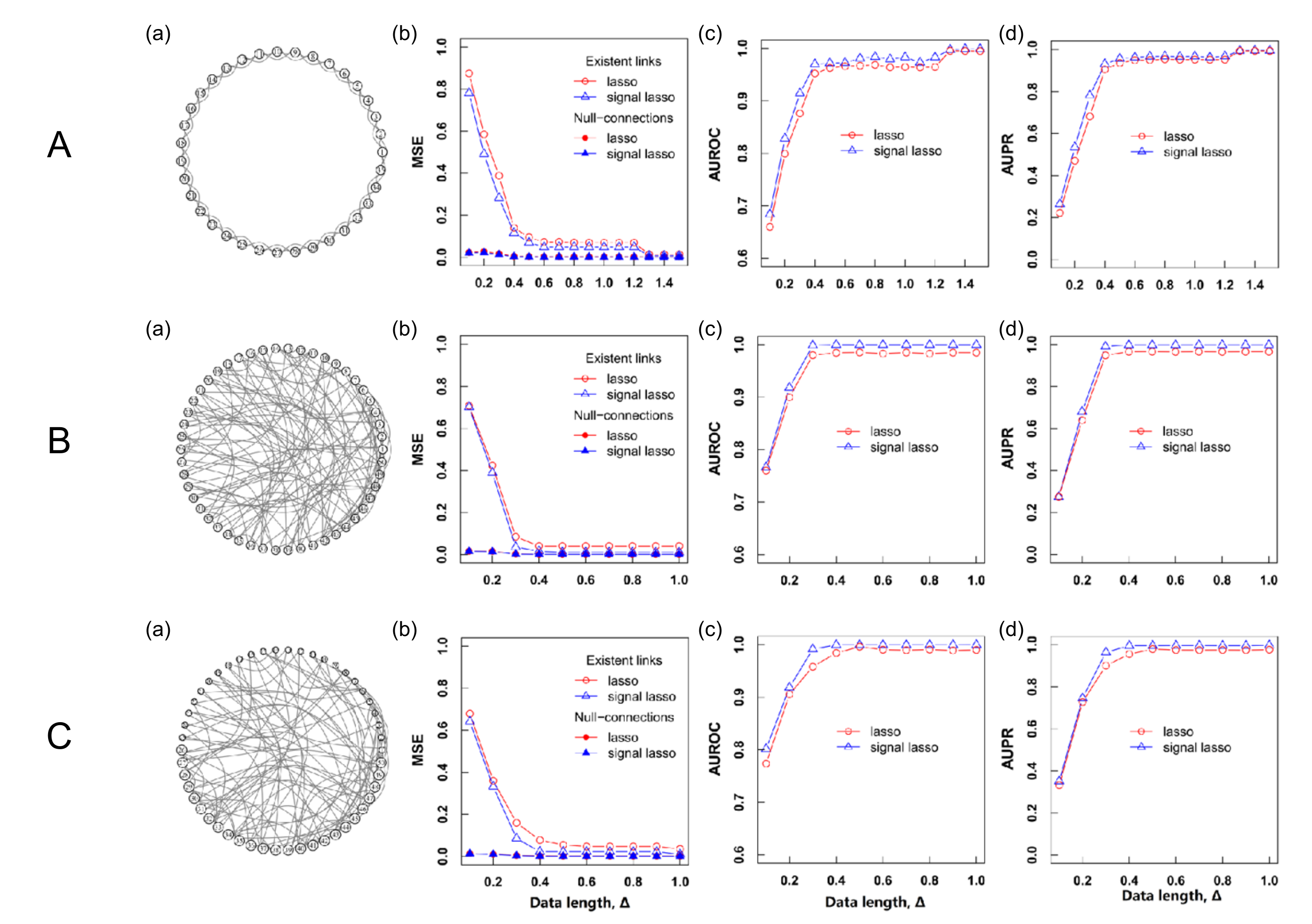}}
	\caption{Panel A refers to the results of the experimental ring network, where Panels A(a) is the scheme of experiment, A(b) to A(d) refer to the MSE criterion, AUROC criterion, and the AUPR criterion, respectively. There are 35 nodes and 140 links and the degree of each node is 4. Panel B refers to the results of the experimental homogeneous random network, where panel B(a) is the scheme of experiment, panels B(b) to B(d) refer to the MSE criterion, AUROC criterion, and the AUPR criterion. There are 50 nodes and 200 links and the degree of each node is 4. Panel C refers to the results of the experimental heterogeneous random network, where panel C(a) is the scheme of experiment, panels C(b) to C(d) refer to the MSE criterion, AUROC criterion, and the AUPR criterion. There are 50 nodes and 200 links and the average degree of each node is 4.
	}
	\label{fig:5}    
\end{figure*}

In this section, we present the results of social network reconstruction using a real data from human behavior experiment~\cite{li2018punishment}, where the purpose is to study the impact of the punishment on network reciprocity. A total of 135 participants from Yunnan University of Finance and Economics and Tianjin University of Finance Economics took part in the experiments and three trials are separately designed and carried out. In treatment I, 35 participants from Tianjin university of finance and economics played iterative prisoner's dilemma game with punishment on the static ring network with four neighbors, where there are 35 nodes and 140 links. For the other two treatments, 100 participants from Yunnan university of finance and economics were invited to participant in the iterated prisoner's dilemma experiment. The treatment II was implemented on the homogeneous random network with degree of 4, where there are 50 nodes and 200 links. While in treatment III, each player was placed on the heterogeneous random network, in which the degree of half nodes is 3 and the degree of other half nodes is 5, there are 50 nodes and 200 links. The network structure is illustrated in Fig. ~\ref{fig:5}A(a), Fig. ~\ref{fig:5}B(a), and Fig. ~\ref{fig:5}C(a), respectively. The employed payoff matrices are given in Table ~\ref{t3}. In each round of the treatment, each player played with its direct neighbors to gain their payoff and updated its strategy to optimize its future payoff. The number of interactions in each session was set to 50, and the number of interactions was undisclosed until the sessions ended. In order to solve the network reconstruction problem in these examples, we thus recorded all the strategies and payoffs generated in the experiment and these information is the available data base for the reconstruction. Through this data base, we can easily construct a linear regression model and obtain the reconstruction results under lasso and signal lasso method. The results are summarized in Fig.\ref{fig:5}(b)-(d). We observe that signal lasso method reaches higher accuracy compared with lasso method. The improvement by using signal lasso is especially obvious for less data set in Fig.\ref{fig:5}A (with relative length of about 0.5 (treatment I)) and Fig.\ref{fig:5}C (with relative length of about 0.4 (treatment III)).

\begin{table}[!htb]
	\caption{textPayoff matrices used in the experiments. The payoff matrix of prisoner's dilemma game with punishment option (treatment I) and the standard prisoner's dilemma (treatment II and treatment III) are shown in left panel and right panel, respectively.}
    \label{t3}
	\begin{minipage}{.5\linewidth}
		\centering
		\setlength{\tabcolsep}{4mm}{
			\begin{tabular}{cccc}
				& C & D & P \\
				\hline
				C & 2 & -2 & -5 \\
				D & 4 & 0 & -3  \\
				P & 2 & -2 & -5  \\
				\hline
		\end{tabular}}
	\end{minipage}%
	\begin{minipage}{.5\linewidth}
		\centering
		\setlength{\tabcolsep}{4mm}{
			\begin{tabular}{ccc}
				~ & C & D  \\
				\hline
				& & \\
				C & 4 & -2  \\
				D & 6 & 0   \\
				\hline
		\end{tabular}}
	\end{minipage}
\end{table}

\section{\label{sec6}Conclusions}
To conclude, we introduced a new method (the signal Lasso) for the estimation problem of signal parameters in the area of network reconstruction. By adding a control term of a L1 norm to shrink the parameters to 1 in the penalty function of Lasso, the estimated signal parameters can be compressed to 0 or 1, which ensures higher reconstruction accuracies compared with Lasso and Compressed Sensing methods. The detailed theoretical properties of the method (including the geometric interpretation and algorithms) are presented in Appendix. For some complicated models with a common feature that the estimated parameters have signal properties, after some modifications such as replacing the least square term by other more suitable measures (for instance -likelihood function), the problem could be solved effectively using idea suggested in this paper.

It is noteworthy that our method is not limited to the two models considered in this letter. For networked binary-state dynamical data~\cite{gleeson2013binary}, including SIS model, majority vote model, and language model etc., some problems can be converted to the models indexed with signal parameters~\cite{li2017universal,shen2014reconstructing,boccaletti2000control}, thus the methods suggested in this paper can be used to obtain more efficient results.
This novel method can be used also with data collected in real-world networks, for gene-regulatory network reconstruction from the expression data in systems biology~\cite{Gardner2003,Hecker2009}, or to infer various functional networks in the human brain ~\cite{Supekar2008}, or for uncovering organizational networks in social and economic science. Due to its robustness against noise and general good performances, our method may have therefore wide applications.  For large scale networks, some dimensional reduction  methods  can be used first to  overcome the dimension curse, and then employ the method suggested here to give the solutions~\cite{shi2020recovering}.

\section{Acknowledgement}
Authors acknowledge the support from National Natural Science Foundation of China (Grants No. 11931015, 11671348, U1803263), the National 1000 Young Talent Plan (No. W099102), and the China Scholarship Council (No. 201908530225).

\clearpage

\appendix{{\bf\Large Appendix}}

\section{Proof of the results in signal Lasso}

We rewrite the above likelihood function in a general matrix form  as
\begin{equation}\label{eqa04}
L(X, \lambda_1,\lambda_2\vert Y,\Phi ) =\frac{1}{2} \Vert Y-\Phi X\Vert^2_2+\lambda_1 \Vert X\Vert_1+\lambda_2\Vert X-{\bf 1}_p\Vert_1,
\end{equation}
where $\lambda_1, \lambda_2 \geq 0$ are two tuning parameters.
$PF(x)=\lambda_1 \Vert x\Vert_1+\lambda_2\Vert x-{\bf 1}_p\Vert_1$ is called penalty function. Fig.\ref{fig:a1}  (a)-(d) list constraint regions $PF(x)=c$ of four different shrink estimation methods, where first three are well known methods: ridge regression~\cite{Hoerl1970ridge}, Lasso estimation and elastic net estimation~\cite{Zou2005elasticnet}. The elastic net method is a balance between ridge regression and Lasso method. Our signal lasso as shown in Fig.\ref{fig:a1} (d) have edges less than or equal to 8 depending on the values of c.

\begin{figure}[htbp!]
	\renewcommand\thefigure{A\arabic{figure}}
	\setcounter{figure}{0}
	\centering{\includegraphics[width=0.45\textwidth]{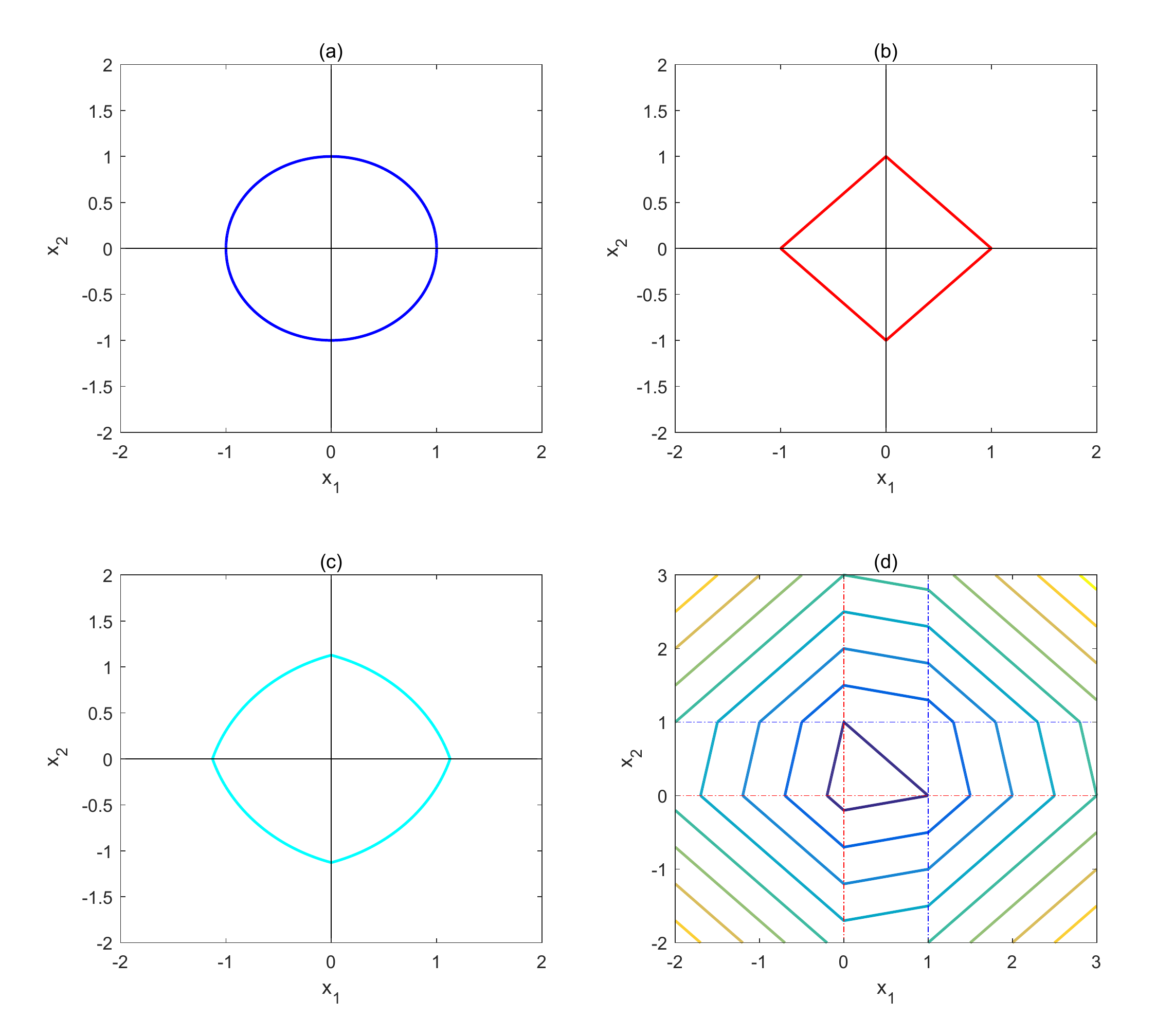}}
	\caption{Constraint regions of $PF(x)=c$ for some constant c under four lasso-type estimation methods in two dimensional case ($p=2$). (a) ridge regression with penalty function $PF(x)=\sum_{j=1}^2 x_j^2$. (b) Lasso estimation with penalty function $PF(x)=\sum_{j=1}^2 \vert x_j \vert $. (c) Elastic net estimation with penalty function $PF(x)=\frac 1 2 \alpha \sum_{j=1}^2 x_j^2+(1-\alpha) \sum_{j=1}^2 \vert x_j\vert$, $\alpha=0.5$. (d) signal lasso with penalty function $PF(x)=\lambda_1 \sum_{j=1}^2 \vert x_j\vert+\lambda_2 \sum_{j=1}^2 \vert x_j-1\vert$ for different values of c (to show the pattern for different c), where $\lambda_1=0.6,\lambda_2=0.4$ }
	\label{fig:a1}
\end{figure}

In order to study the geometry of signal Lasso, we assume that the columns of $\Phi$ are orthogonal each other and $p < n$. The ordinary least squares estimate in this special case then has the form of $\hat X_0=\Phi^\top Y$. Let $\hat{Y}_0=\Phi \hat X_0$, we then have
\begin{equation}\label{eqa05}
\begin{array}{ll}
L(X, \lambda_1,\lambda_2\vert Y,\Phi )
=&\frac{1}{2} \Vert Y-\hat Y_0\Vert^2_2+\frac{1}{2} \Vert X-\hat X_0\Vert^2_2+\\
& \lambda_1 \Vert X\Vert_1+\lambda_2\Vert X-\bf{1}_p\Vert_1.
\end{array}
\end{equation}
Note that the first term in the right is constant with respect to X and $\Vert \Phi (X-\hat X_0)\Vert_2^2=\Vert (X-\hat X_0)\Vert_2^2$. Using the fact ~\cite{hastie2015book} that
\begin{equation}\label{eqa07}
\begin{array}{ll}
\displaystyle\frac{\partial \vert x \vert}{\partial x}&=\displaystyle\frac{\partial (\mbox{sign}(x) x )}{\partial x} \\
&=\left\{
\begin{array}{l}
1, \ \ \ \ \ \ \ if \ x>0, \\
-1, \ \ \ \ \ \ \ if \ x < 0, \\
\mbox{some} \ \mbox{values in} \ [-1,1] \ \ if \ x=0,
\end{array}\right.
\end{array}
\end{equation}
and differentiating $L(X, \lambda_1,\lambda_2\vert Y,\Phi )$ with respect to $X$
and set it to zero, after some calculations we have
\begin{equation}\label{eqa08}
\begin{split}
\hat X_k=\left\{
\begin{array}{l}
(\hat X_{k0}+\delta_1)_{-}, \ \  \hat X_{k0}\leq0, \\
(\hat X_{k0}-\delta_2)_{+}, 0<\hat X_{k0}\leq 1+\delta_2, \\
\max\{1, \hat X_{k0}-\delta_1\}, \hat X_{k0}> 1+\delta_2,
\end{array}\right.
\end{split}
\end{equation}
for $k=1,\cdots, p$, where $\delta_1=\lambda_1+\lambda_2$ and  $\delta_2=\lambda_1-\lambda_2$, $\hat X_{k0}$ and $\hat X_k$ are the $k$th element of $\hat X_{0}$ and $\hat X$, respectively; $B_{+}$ denote the positive part of $B$, it means that $B_{+}=B$ if $B\ge 0$ and 0 otherwise. $B_{-}$ is similarly defined as the negative part of $B$.

If a intercept need to be included in the model, our method is also available. Since the intercept will not be shrunk we simply center the covariates $\phi_{ij}$, and then the intercept can be replaced by $\bar{y}=\sum_{i=1}^n y_i/n$. Having solved for this optimal estimator, it remains to compute estimators of other coefficient using signal Lasso.

\section{ Algorithm and computation}

It is noted that penalty function $PF(x)=\lambda_1 \sum_{j=1}^p \vert x_j\vert+\lambda_2 \sum_{j=1}^p \vert x_j-1\vert$ is a convex function. Thus the optimization problem in (\ref{eq4}) does not suffer from the multiple local minimal
issue, and its global minimizer can be efficiently solved.  A variety of algorithm are available in some software with package of convex optimization program and can be used to calculate the estimation of $X$ in signal Lasso method.

However we can provide an alternative algorithm for signal Lasso using the coordinate descent method (~\cite{hastie2015book}, pp109-118) as that in Lasso estimation method. Below we give more detail:

Coordinate descent method is an iterative algorithm that update the estimator by choosing a single coordinate to update and then performing a univariate minimization over this coordinate. Differentiating (\ref{eq4}) with respect to $X_k$ and equal to zero, we have
\begin{equation}\label{eqa09}
-\sum_{i=1}^n (y_i-\sum_{j=1}^p \phi_{ij}X_j)\phi_{ik}+\lambda_1 s_k^{(1)} + \lambda_2 s_k^{(2)}=0, \\
\end{equation}
where $s_k^{(1)}=\partial \sum_{j=1}^p\vert X_k \vert/\partial X_k$, $s_k^{(2)}=\partial \sum_{j=1}^p\vert X_k -1\vert/\partial X_k$. Let $r^{(k)}=\sum_{i=1}^n (y_i-\sum_{j\neq k} \phi_{ij}X_j)$ denote the partial residual, then using formula (\ref{eqa07}) and after some calculations, we have $X_k=$
\begin{equation}\label{eqa10}
\left\{
\begin{array}{l}
\displaystyle\bigg[\frac{\left< r^{(k)}, \phi_k\right>}{\left< \phi_k, \phi_k\right>}+\delta_{1k}^*\bigg]_{-}, \ \ \ \ \ \ \ \frac{\left< r^{(k)}, \phi_k\right>}{\left< \phi_k, \phi_k\right>}\leq 0, \\
\displaystyle\bigg[\frac{\left< r^{(k)}, \phi_k\right>}{\left< \phi_k, \phi_k\right>}-\delta_{2k}^*\bigg]_{+}, \ \ \ \ \ \ \ 0<\frac{\left< r^{(k)}, \phi_k\right>}{\left< \phi_k, \phi_k\right>} \ \leq 1+\delta_{2k}^*, \\
\displaystyle\max\left\{1, \bigg[\frac{\left< r^{(k)}, \phi_k\right>}{\left< \phi_k, \phi_k\right>}-\delta_{1k}^*\bigg]_{+}\right\}, \ \ \frac{\left< r^{(k)}, \phi_k\right>}{\left< \phi_k, \phi_k\right>} \ > 1+\delta_{2k}^*,
\end{array}\right.
\end{equation}
where $\left< z, y\right>$ denote the inner product of vectors $z$ and $y$, $\delta_{1k}^*=(\lambda_{1k}+\lambda_{2k})/\left< \phi_k, \phi_k\right>$ and  $\delta_{2k}^*=(\lambda_{1k}-\lambda_{2k})/\left< \phi_k, \phi_k\right>$. From the definition of $S_{\theta_1, \theta_2}(z)$, it is easy to see
\begin{equation}
X_k=S_{\delta_{1k}^*, \delta_{2k}^*}\left( \displaystyle\frac{\left< r^{(k)}, \phi_k\right>}{\left< \phi_k, \phi_k\right>}\right),
\end{equation}
where threshold function $S_{\theta_1,\theta_2}(z) $ is defined by
\begin{equation}\label{eqa11}
S_{\theta_1,\theta_2}(z)=\left\{
\begin{array}{l}
(z+\theta_1)_{-},  \qquad  z \leq 0,\\
(z-\theta_2)_{+}, \qquad  0< z \leq 1+\theta_2,\\
\max\{1, z-\theta_1 \}, \qquad  z > 1+\theta_2.
\end{array}\right.
\end{equation}
Note that
$\left<r^{(k)}, \phi_k\right>=\left< r, \phi_k\right>+X_k \left< \phi_k, \phi_k\right>$,  where $r=Y-\Phi X$, we have
\begin{equation}\label{eqa12}
X_k=S_{\delta_{1k}^*,\delta_{2k}^*}\left( X_k+\displaystyle\frac{\left< r, \phi_k\right>}{\left< \phi_k, \phi_k\right>}\right)
\end{equation}
Therefore the update can be written as
\begin{equation}\label{eqa13}
\hat X_k^{t+1}\leftarrow S_{\delta_{1k}^*,\delta_{2k}^*}\left(\hat X_k^t+\displaystyle\frac{\left< \hat r^t, \phi_k\right>}{\left< \phi_k, \phi_k\right>} \right)
\end{equation}
where $\hat X_k^{t}$ denote the estimator of $X_k$ in the $t$th step, and $\hat r^t=Y-\Phi \hat X^t$. The overall algorithm operates by applying this update repeatedly in a cyclical manner, updating the coordinates of $\hat X$ along the way. Once an initial estimator of $X$ is given, for example by Lasso estimation or ridge estimation, the update can be continued until convergence. The detailed algorithm based on coordinate descent method is given in Algorithm \ref{alg:1}.
\begin{algorithm}[H]
	\caption{Algorithm for computing signal Lasso estimation using coordinate descent method.}
	\label{alg:1}
	\begin{algorithmic}[1]
        \Require ~\\
        The response vector $Y$ and the design matrix $\Phi$.
        \Ensure ~\\
        Estimated signal parameter vector $X$ giving $Y$ and $\Phi$.
		\State {\textbf{Step 1: Iteration over all coordinator of $X$}: Assignment of an initial value for parameter $X$, which can be lasso estimator or ridge estimator, denoted by  $\hat{X}^{(0)}$.}
		\State {\textbf{Step 2} Set an initial threshold $\varepsilon=1$.}
		\Repeat
		\State{\textbf{Step 2.1: Iteration for each coordinator $X_k$}.}~\\
		\textbf{for $k=1,\cdots, p$  ~~ do} \\
		Set an initial threshold: $\varepsilon_k=1$, and set $t=0$
		\Repeat
		\State {\textbf{\emph{Step 2.1.1}}. Calculate the solution $\hat{X}_{k}^{(t+1)}$ by updating formula Eq. (\ref{eqa13}).}
		\State {\textbf{\emph{Step 2.1.2}}. Calculate the estimating precision for estimating $X_k$: $\varepsilon_k=\vert \hat{X}_k^{(t+1)}-\hat{X}_k^{(t)} \vert$. Set $t=t+1$.}
		\Until{$\varepsilon_k< 1e-8$.} \\
		Updating the value of $\hat X_{k}^{(t)}$ using obtained new estimator by $\hat X_{k}^{(t)}=\hat X_{k}^{(t+1)}$. \\
		\textbf{end for} \\
		Obtain a new estimator  $\hat X$ from converged coordinator.
		\State {\textbf{Step 2.2}.   Calculate the estimating precision for estimating $X$: $\varepsilon=\Vert \hat{X}-\hat{X}^{(0)}\Vert_1 =\sum_{k=1}^p \vert \hat{X}_k-\hat{X}_k^{(0)}\vert$. Set $\hat{X}^{(0)}=\hat{X}$ }
		\Until{$\varepsilon< 1e-10$.}
		\State{\textbf{Step 3} Output the final estimator of $\hat X$. }
	\end{algorithmic}
\end{algorithm}
\section{The choice of tuning parameter}

One of penalty function as we mentioned in (\ref{eq2}) is given by
\begin{equation}\label{eqa14}
PF(X, \lambda,\alpha)=\lambda_1 \Vert X\Vert_1+\lambda_2 \Vert X-{\bf 1}_p\Vert_1,
\end{equation}
The use of $(\lambda,\alpha)$ has no much differences with use of $(\lambda_1,\lambda_2)$, but latter is more convenient for flexible expression. Another reparameterization of tuning parameters are that based on $(\delta_1,\delta_2)$ as we shown in (\ref{eqa08}).
Obviously there is a one-by-one transformation between $(\lambda_1,\lambda_2)$ and $(\delta_1,\delta_2)$. The reparameterization based on $(\delta_1,\delta_2)$ is more convenient as shown in Fig.~\ref{fig:1}(b)-(c) in the paper. The $\delta_2$ should not be too large since the signal Lasso will shrink the $\hat X_{k0}$ that larger than $1+\delta_2$ (and less than $1+\delta_1$) toward to 1. As a result it will shrink the $\hat X_{k0}<1+\delta_2$ (and larger than $\delta_2$) to a less value, which means value of $\hat X_0$ will be shrink to $1-\delta_2$ as shown in  Fig.~\ref{fig:1}(b)-(c). A reasonable range for $\delta_2$ should be in interval $(0.1,0.4)$. The choice of $\delta_1$ will assure that some larger values such as $1+\delta_2\le\hat X_{k0}\le 1+\delta_1$ shrink to 1, and some small values such that $-\delta_1\le\hat X_{k0}\le \delta_2$ shrink to 0 depending on the situations. In real example and simulations, we find some estimators have more negative values  when there are noises in the data, the larger $\delta_1$ is required in these special cases.

Similarly we also can use the expression
\begin{equation}\label{eqa15}
PF(X, \lambda,\alpha)=\lambda [(1-\alpha) \Vert X\Vert_1+\alpha \Vert X-{\bf 1}_p\Vert_1 ],
\end{equation}
to find tuning parameters $\lambda$ and $\alpha$. It also involve the use of cross-validation of two parameters.

The choices of tuning parameters $(\lambda_1,\lambda_2)$ or $(\delta_1,\delta_2)$ or $(\lambda,\alpha)$ can be conducted using well known cross-validation (CV) technique~\cite{hastie2015book}. In more detail, firstly the full data set is divided into some number of groups $K>1$ ( called by K-fold). The typical choices of $K$ might be 5 or 10. If $K=n$, it corresponds to leave-one-out test. We fix one group as the test set, and remaining $K-1$ group are designed as training set. We then apply signal Lasso method based on training set to obtain a fitted model for a range of $(\lambda_1,\lambda_2)$, and use each fitted model to predict the responses in test set, which lead to a mean square prediction error (MSPE) for each combination of $(\lambda_1,\lambda_2)$. This procedure can be repeated $K$ times and average MSPE is recorded. The minimum value corresponds to some combination of $(\lambda_1,\lambda_2)$ is selected as best choice of tuning parameters in our method. Because two tuning parameters are involved in signal Lasso, we have to search a best combination of
$(\lambda_1,\lambda_2)$ in a two-dimensional surface for possible ranges of $\lambda_1$ and $\lambda_2$.

In our simulation studies based on evolutionary game scheme, the observations are measured as a time series from 1 to $L$. Thus we use the observations in time $L$ as test set, and the first $L-1$ observations as training set. MSPE can be defined as
\begin{equation}\label{eqa16}
MSPE(\lambda_1, \lambda_2)=\frac{1}{N} \sum_{i=1}^N (\hat y^{-L}_{iL}-y_{iL})^2,
\end{equation}
where $N$ is the size of network, $\hat y^{-L}_{iL}$ is a prediction of $L$th observation using first $L-1$ observations in Eq. (\ref{eq9}) of the main manuscript. Best choice of $(\lambda_1,\lambda_2)$ will be the one that having the smallest $MSPE$.

\bibliographystyle{iopart-num}

\bibliography{bibio}

\clearpage
\onecolumngrid
\setcounter{equation}{0}
\renewcommand\theequation{S\arabic{equation}}
\setcounter{page}{1}
\renewcommand\thepage{S\arabic{page}}

\section*{
	Supplementary Information for\\
	``Inferring Network Structures via Signal Lasso''}

We provide more details for the results presented in Table.~\ref{t1} in the main text. Fig.\ref{fig:s2}-\ref{fig:s5} show that in the absence of noise, results of mean square error as a function of data amount $\Delta$ for compressed sensing (CS), lasso, and signal lasso from three representative networks based on spatial ultimatum games. The employed theoretical networks with size $N=100$ are: BA scale-free networks (left), WS small-world networks (middle), and ER random networks (right). For these networks with different average degree, we can observe that, on one hand, the proposed method has better performance than lasso and CS and this advantage becomes more obvious with the increases of network average degree. On the other hand, there is no significant difference between the results of CS and lasso. In the presence of noise, as shown in Fig.\ref{fig:s6} and Fig.\ref{fig:s7}, signal lasso can still performed better than lasso, however, this advantage is less obvious especially for large noise variance (Fig.\ref{fig:s7}). CS, a method for dealing with the problems of sparse signal reconstruction, is frequently used in this problem and its effectiveness has been validated. However, this method is sensitive to noise for that its condition of convex optimization problem $Y = \Phi X$ can not be guaranteed when faced with noise. In our situations, the solutions of CS is unstable and we cannot even obtain its solutions if the Gaussian noise $N (0, \sigma^2)$ is add to the observation vector $Y$. We have also tested the performance of signal lasso to several empirical networks (Fig.\ref{fig:s8}), and get the consistent results with theoretical networks. In a word, the proposed signal lasso method is effective for problems of signal parameters and it has strong robustness against noise.

\clearpage

\renewcommand\thefigure{S\arabic{figure}}
\setcounter{figure}{0}  
\section*{Supplementary Figures}

\begin{figure*}[htbp!]
	\renewcommand\thefigure{S\arabic{figure}}
	\centering{\includegraphics[width=1.01\textwidth]{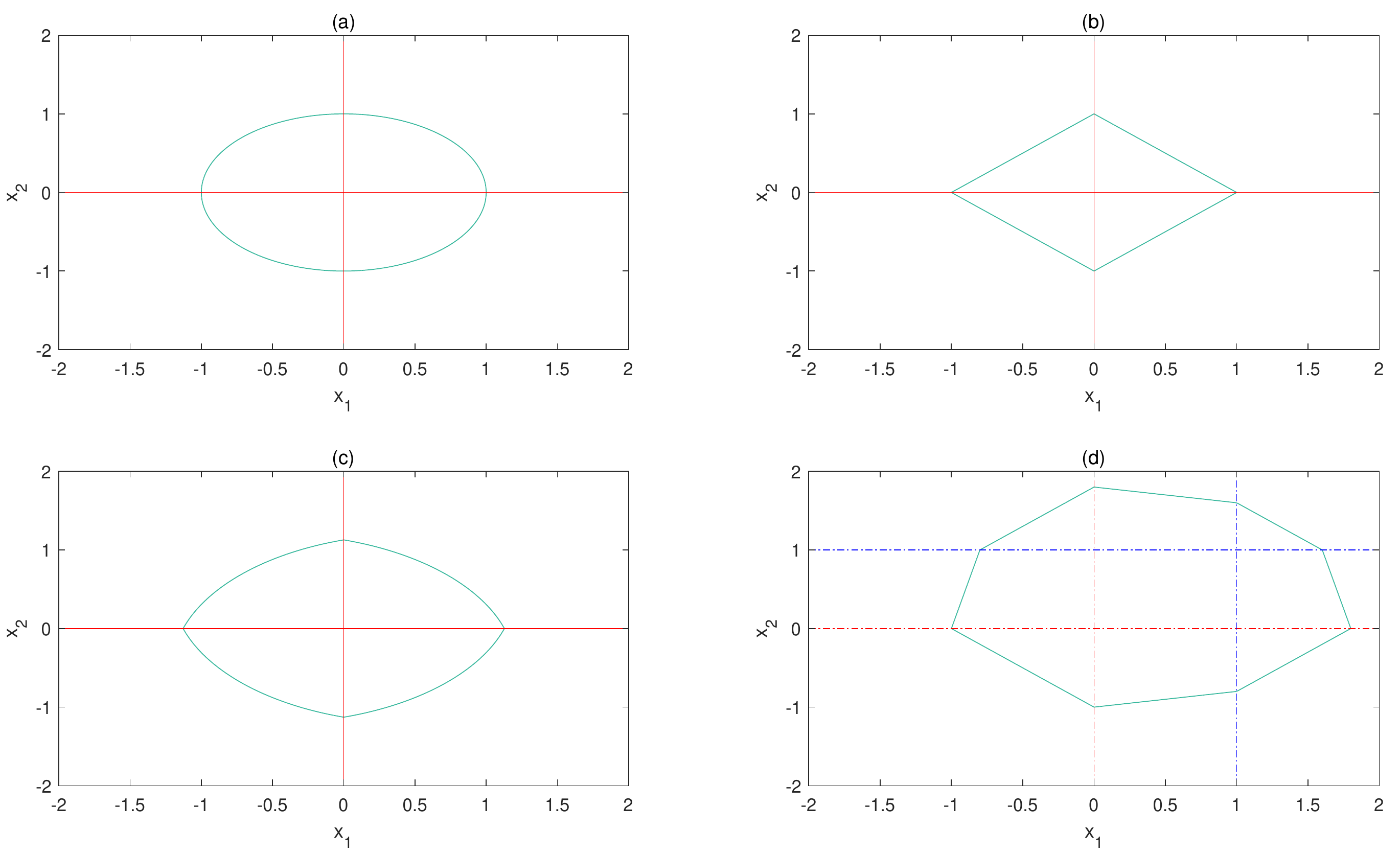}}
	\caption{Constraint regions of $PF(x)=c$ for some constant c under four lasso-type estimation methods in two dimensional case ($p=2$). (a) ridge regression with penalty function $PF(x)=\sum_{j=1}^2 x_j^2$. (b) Lasso estimation with penalty function $PF(x)=\sum_{j=1}^2 \vert x_j \vert $. (c) Elastic net estimation with penalty function $PF(x)=\frac 1 2 \alpha \sum_{j=1}^2 x_j^2+(1-\alpha) \sum_{j=1}^2 \vert x_j\vert$, $\alpha=0.5$. (d) signal lasso with penalty function $PF(x)=\lambda_1 \sum_{j=1}^2 \vert x_j\vert+\lambda_2 \sum_{j=1}^2 \vert x_j-1\vert$, $\lambda_1=0.6,\lambda_2=0.4$ }
	\label{fig:s1}
\end{figure*}
\vfill
\clearpage
\vfill
\begin{figure*}[htbp!]
	\renewcommand\thefigure{S\arabic{figure}}
	\centering{\includegraphics[width=0.99\textwidth]{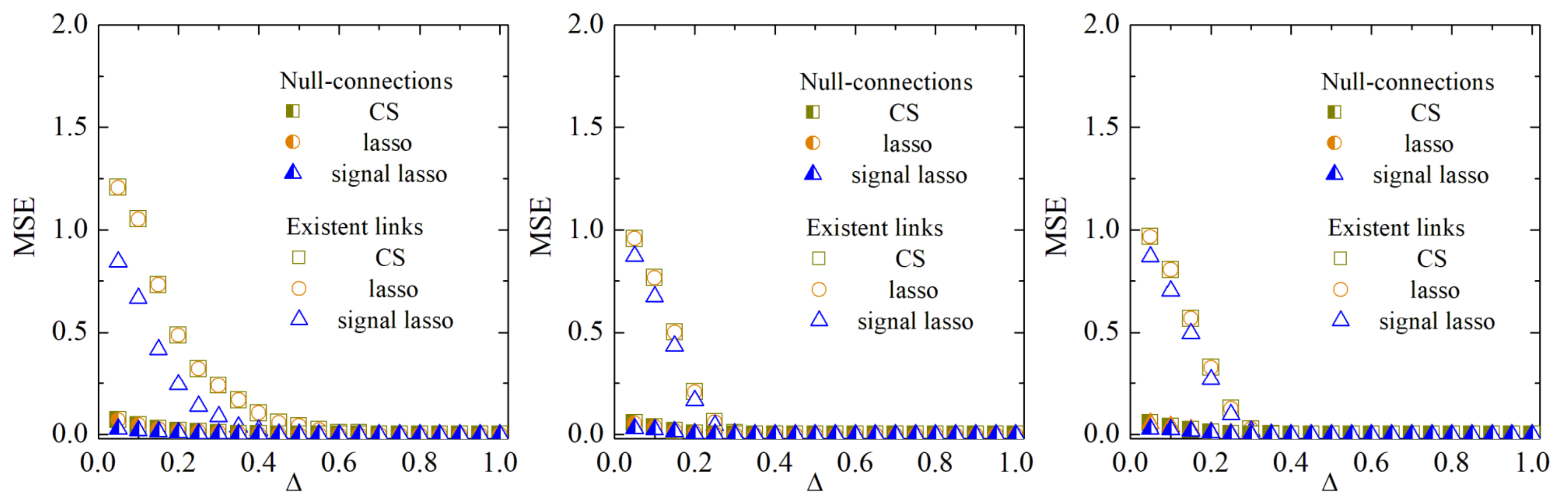}}
	\caption{Results of mean square error (MSE) of existent links and null connections with compressed sensing (CS), lasso, and signal lasso from BA scale-free networks (left), WS small-world networks (middle), and ER random networks (right) based on evolutionary spatial ultimatum games. Each data point is obtained by averaging over ten independent runs. Network size $N$ is 100, average degree $\left \langle k \right \rangle = 6$.}
	\label{fig:s2}       
\end{figure*}
\vfill
\clearpage
\vfill
\begin{figure*}[htbp!]
	\renewcommand\thefigure{S\arabic{figure}}
	\centering{\includegraphics[width=0.99\textwidth]{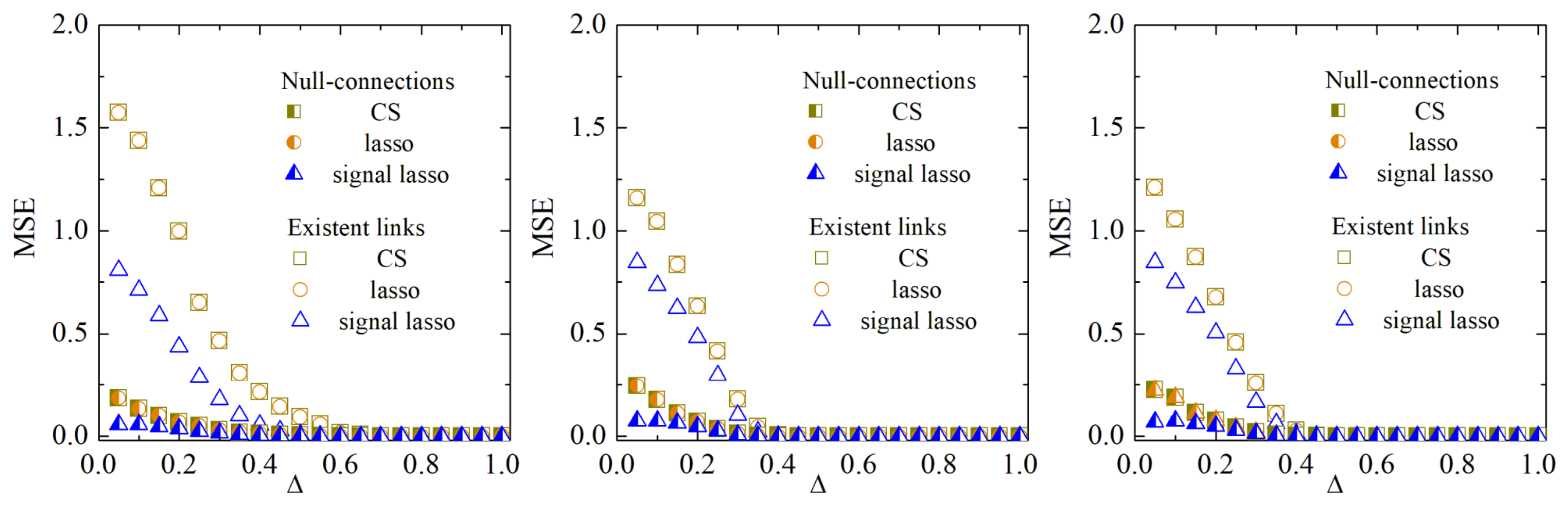}}
	\caption{Results of mean square error (MSE) of existent links and null connections with compressed sensing (CS), lasso, and signal lasso from BA scale-free networks (left), WS small-world networks (middle), and ER random networks (right) based on evolutionary spatial ultimatum games. Each data point is obtained by averaging over ten independent runs. Network size $N$ is 100, average degree $\left \langle k \right \rangle = 12$.}
	\label{fig:s3}       
\end{figure*}
\vfill
\clearpage
\vfill
\begin{figure*}[htbp!]
	\renewcommand\thefigure{S\arabic{figure}}
	\centering{\includegraphics[width=0.99\textwidth]{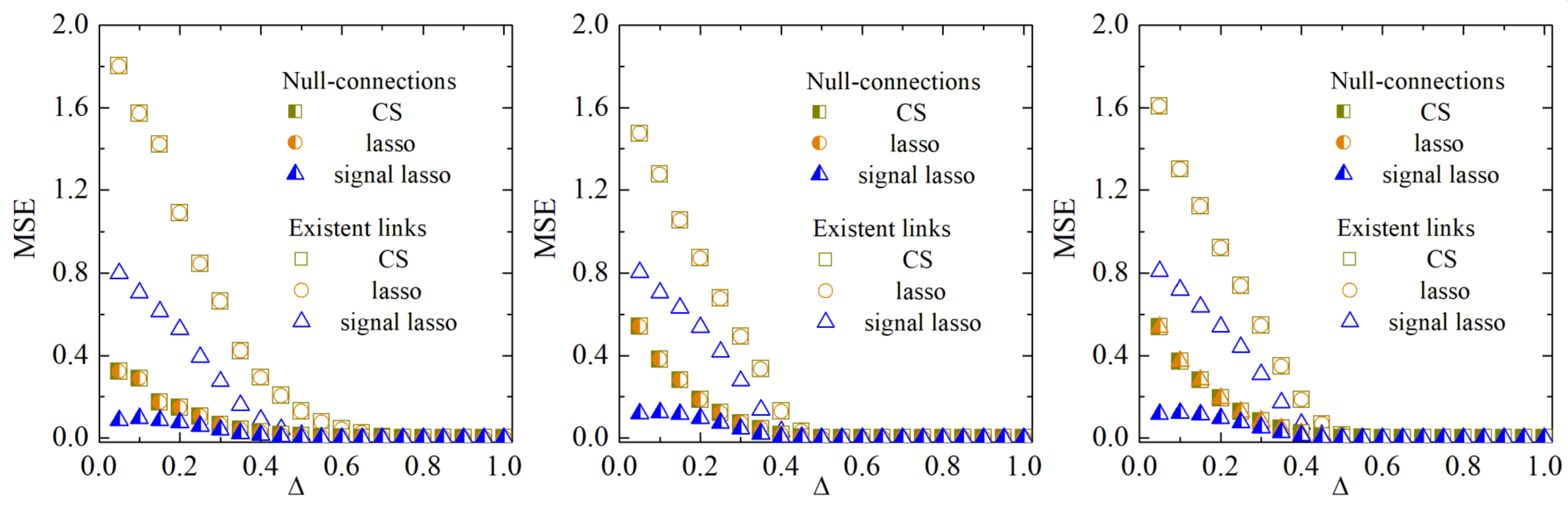}}
	\caption{Results of mean square error (MSE) of existent links and null connections with compressed sensing (CS), lasso, and signal lasso from BA scale-free networks (left), WS small-world networks (middle), and ER random networks (right) based on evolutionary spatial ultimatum games. Each data point is obtained by averaging over ten independent runs. Network size $N$ is 100, average degree $\left \langle k \right \rangle = 18$.}
	\label{fig:s4}       
\end{figure*}
\vfill
\clearpage
\vfill
\begin{figure*}[htbp!]
	\renewcommand\thefigure{S\arabic{figure}}
	\centering{\includegraphics[width=0.99\textwidth]{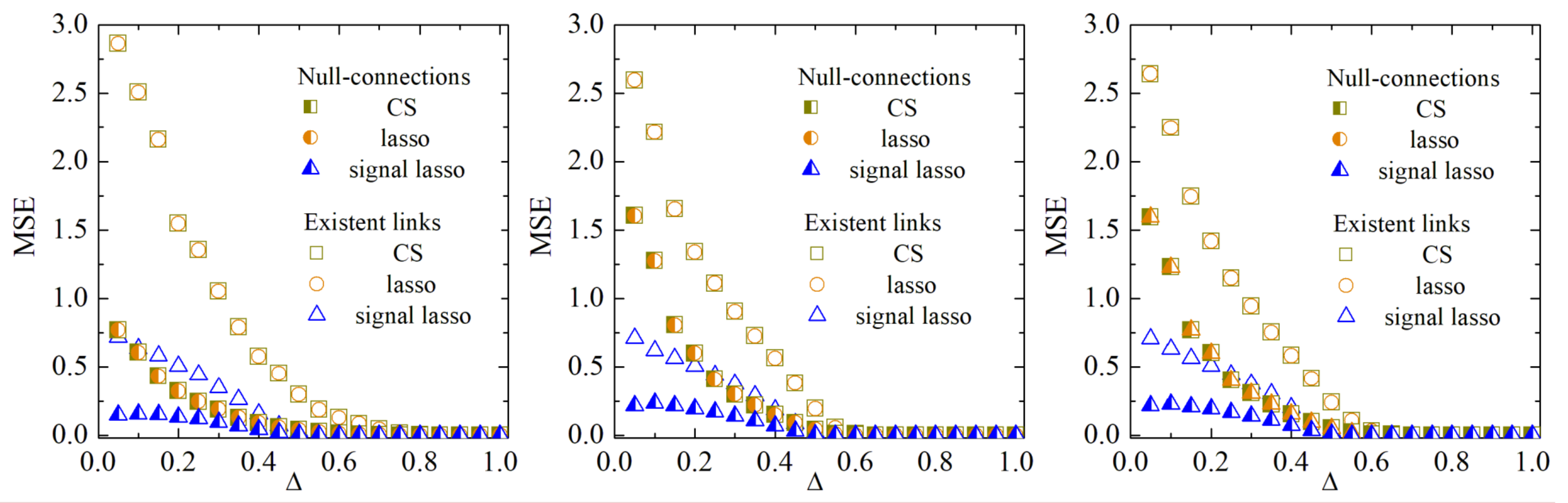}}
	\caption{Results of mean square error (MSE) of existent links and null connections with compressed sensing (CS), lasso, and signal lasso from BA scale-free networks (left), WS small-world networks (middle), and ER random networks (right) based on evolutionary spatial ultimatum games. Each data point is obtained by averaging over ten independent runs. Network size $N$ is 100, average degree $\left \langle k \right \rangle = 30$.}
	\label{fig:s5}       
\end{figure*}
\vfill
\clearpage
\vfill
\begin{figure*}[htbp!]
	\renewcommand\thefigure{S\arabic{figure}}
	\centering{\includegraphics[width=0.99\textwidth]{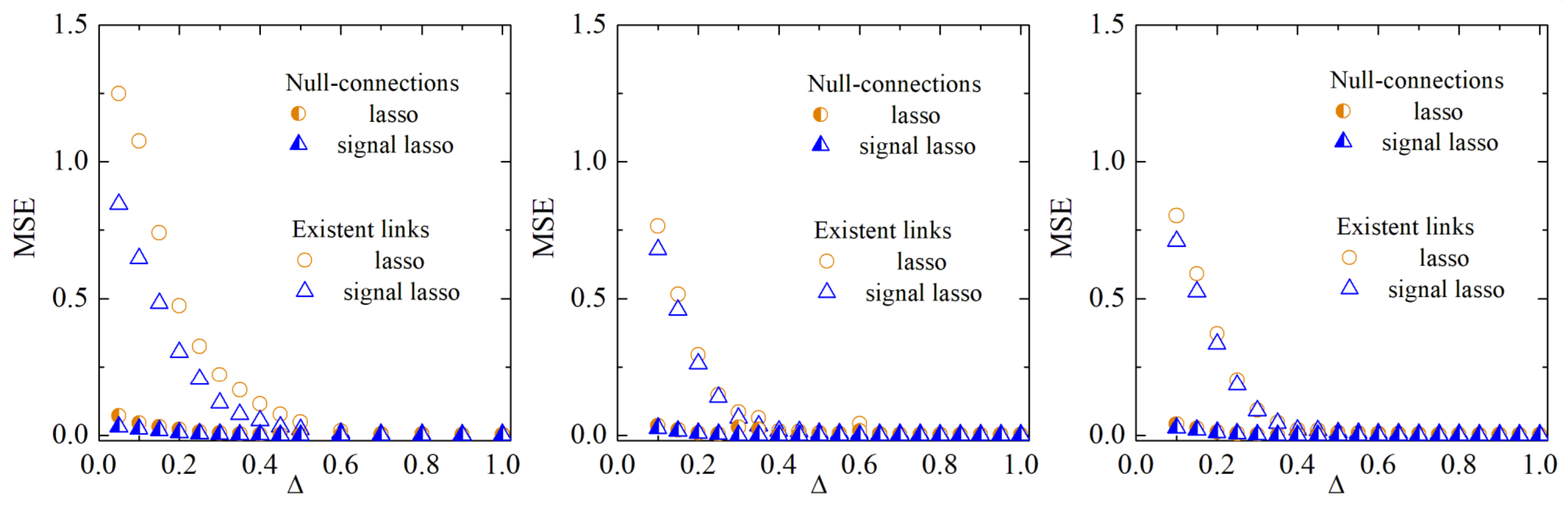}}
	\caption{Results of mean square error (MSE) of existent links and null connections with compressed sensing (CS), lasso, and signal lasso from BA scale-free networks (left), WS small-world networks (middle), and ER random networks (right) based on evolutionary spatial ultimatum games. Each data point is obtained by averaging over ten independent runs. Network size $N$ is 100, average degree $\left \langle k \right \rangle = 6$, a gaussian noise with $N$ (0, $0.05^2$) is implemented to $Y$. }
	\label{fig:s6}       
\end{figure*}
\vfill
\clearpage
\vfill
\begin{figure*}[htbp!]
	\renewcommand\thefigure{S\arabic{figure}}
	\centering{\includegraphics[width=0.99\textwidth]{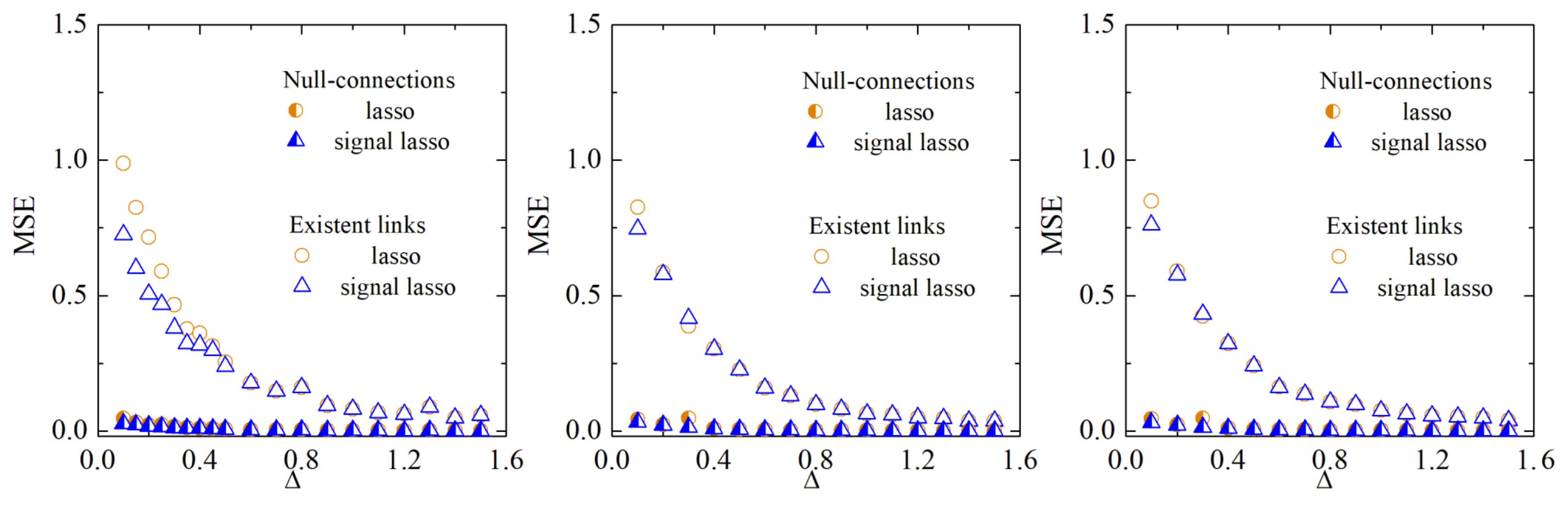}}
	\caption{Results of mean square error (MSE) of existent links and null connections with compressed sensing (CS), lasso, and signal lasso from BA scale-free networks (left), WS small-world networks (middle), and ER random networks (right) based on evolutionary spatial ultimatum games. Each data point is obtained by averaging over ten independent runs. Network size $N$ is 100, average degree $\left \langle k \right \rangle = 6$, a gaussian noise with $N$ (0, $0.3^2$) is implemented to $Y$.}
	\label{fig:s7}       
\end{figure*}
\vfill
\clearpage
\vfill
\begin{figure*}[htbp!]
	\renewcommand\thefigure{S\arabic{figure}}
	\centering{\includegraphics[width=0.99\textwidth]{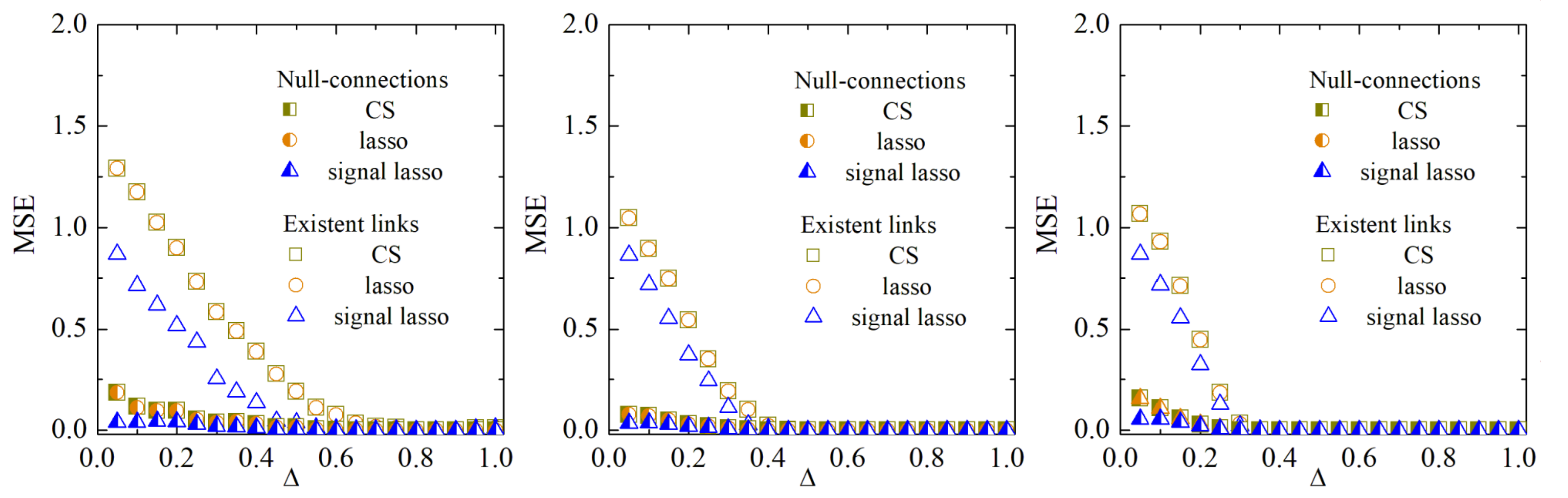}}
	\caption{Results of mean square error (MSE) of existent links and null connections with compressed sensing (CS), lasso, and signal lasso from several empirical networks based on evolutionary spatial ultimatum games. From left to right, the employed networks are: karate, dolphins and football networks, respectively. Each data point is obtained by averaging over ten independent runs.}       
	\label{fig:s8}
\end{figure*}
\vfill
	
\end{document}